\documentclass[english,prb,showpacs]{revtex4}
\usepackage[T1]{fontenc}
\usepackage[latin1]{inputenc}

\makeatletter

\providecommand{\tabularnewline}{\\}

\usepackage{babel}
\makeatother
\begin{document}

\title{Ab initio Wannier-function-based many-body approach to Born charge
of crystalline insulators }

\author{Priya Sony and Alok Shukla}

\affiliation{Physics Department, Indian Institute of Technology, Powai, Mumbai
400076, INDIA}

\begin{abstract}
In this paper we present an approach aimed at performing many-body
calculations of Born-effective charges of crystalline insulators,
by including the electron-correlation effects. The scheme is implemented
entirely in the real space, using Wannier-functions as single-particle
orbitals. Correlation effects are computed by including virtual excitations
from the Hartree-Fock mean field, and the excitations are organized
as per a Bethe-Goldstone-like many-body hierarchy. The results of
our calculations suggest that the approach presented here is promising. 
\end{abstract}

\pacs{77.22.-d, 71.10.-w, 71.15.-m}

\maketitle
The Born effective charge (BEC) of a periodic solid is an important
phenomenological quantity which connects the electronic structure
of the system to its phononic properties.\cite{born} Of late, in
the context of ferroelectric materials and their phase transitions,
BEC has generated tremendous amount of interest.\cite{zhong} Using
BEC, one can also describe the lattice dynamics, and its coupling
to infrared radiation, in a simple intuitive manner.\cite{car} Most
of the modern calculations of BECs are based upon the Berry-phase-based
theory of macroscopic polarization developed by King-Smith and Vanderbilt.\cite{vanderbilt}
The aforesaid formalism is based upon single-particle orbitals, and
therefore, can be implemented in a straight-forward manner within
\emph{ab initio} density-functional theory (DFT),\cite{zhong,car}
or the Hartree-Fock (HF) framework.\cite{dall} As far as many-body
calculations of polarization properties are concerned, Martin and
coworkers have proposed several approaches which, to the best of our
knowledge, have not been implemented within an \emph{ab initio} methodology.\cite{martin-1,martin-2}
Filippetti and Spaldin have recently implemented an \emph{ab initio}
method aimed at including correlation effects by using a self-interaction-corrected
(SIC) density-functional approach\cite{filip}. 

Recently, we have developed a wave-function-based \emph{ab initio}
methodology aimed at performing electronic structure calculations
on crystalline insulators.\cite{shukla1,shukla2,shukla4,shukla5}
The approach uses Wannier-functions as single particle orbitals obtained
at the Hartree-Fock level, which can subsequently be used to include
the electron correlation effects, if needed. The approach has been
applied to calculate ground state geometries, cohesive energies, and
elastic properties of crystalline insulators at the Hartree-Fock level,\cite{shukla1,shukla2}
as well as at the correlated level.\cite{shukla4,shukla5} Moreover,
within the Berry-phase formalism of King-Smith and Vanderbilt,\cite{vanderbilt}
we have also used our approach to compute the BECs of several ionic
insulators at the HF level.\cite{shukla6} The purpose behind the
present work is to use our Wannier-function-based methodology to perform
correlated calculations of the BEC's of insulators. Since, ours is
a real-space approach, we start with the following expression for
the electronic contribution to the polarization per unit cell (${\bf P}_{e}$)
valid for insulators\cite{note}

\begin{equation}
{\bf P_{e}^{(\lambda)}}=\frac{q_{e}}{N\Omega}\langle\Psi_{0}^{(\lambda)}|{\bf R}_{e}|\Psi_{0}^{(\lambda)}\rangle,\label{eq:corr}\end{equation}

where $\lambda$ is a parameter governing the state of crystal (for
the present case, it represents atomic displacements), $\Omega$ is
the volume of the unit cell, $q_{e}$ is the electronic charge, $N\:(\rightarrow\infty)$,
represents the total number of unit cells in the crystal, ${\bf R}_{e}=\sum_{k=1}^{N_{e}}{\bf r_{k}}$
is the many-particle position operator for the $N_{e}$ electrons
of the crystal, and $|\Psi_{0}^{(\lambda)}\rangle$ represents the
correlated ground-state wave-function of the infinite solid. Next
we verify that for an infinite crystal, Eq. (\ref{eq:corr}) above
is consistent with the Berry-phase-based expression for the BEC's
derived by King-Smith and Vanderbilt\cite{vanderbilt} at the mean-field
level. If we express the \emph{mean-field} (HF or otherwise) ground-state
many-particle wave function $|\Phi_{0}^{(\lambda)}\rangle$of a crystal
in terms of Wannier functions expressed in terms of square-integrable
occupied Wannier functions $\{ W_{n}({\bf r}-{\bf R_{i}}),\: i=1,\ldots,N,\: n=1,\ldots,M\}$
located in the $N$ unit cells constituting the solid,\cite{shukla2}
then using the Slater-Condon rules governing the matrix elements of
a one-body operator between two many-particle states,\cite{mcweeny}
we obtain \begin{equation}
\langle\Phi_{0}^{(\lambda)}|{\bf R}_{e}|\Phi_{0}^{(\lambda)}\rangle=N\sum_{n=1}^{M}f_{n}\int{\bf r}|W_{n}({\bf r})|^{2}d{\bf r}+n_{e}(\sum_{i=1}^{N}{\bf R}_{i}),\label{eq:int}\end{equation}
where ${\bf \{ R_{i},}\: i=1,\ldots,N\}$ are the lattice vectors
of $N$ unit cells of the crystal, $W_{n}(r)$ is the $n$-th Wannier
function of the unit cell, $f_{n}$ is the number of electrons in
the $n$-th Wannier function ($f_{n}=2$, for band insulators), $M$
is the total number of occupied Wannier functions per unit cell, and
$n_{e}=\sum_{n=1}^{M}f_{n}$ is the total number of electrons per
unit cell. However, if the $N$ unit cells are distributed among complete
shells (stars), then $\sum_{i=1}^{N}{\bf R}_{i}=0$. Combining this
result with Eqs. (\ref{eq:int}) and (\ref{eq:corr}), we obtain the
mean-field expression (${\bf P}_{0}^{(\lambda)}$) for the polarization
per cell for a crystal

\begin{eqnarray}
{\bf P}_{0}^{(\lambda)} & = & q_{e}/\Omega\sum_{n=1}^{M}f_{n}\int{\bf r}|W_{n}^{(\lambda)}({\bf r)|^{2}{\bf dr}}.\label{eq:pol}\end{eqnarray}
 This equation is nothing but the Wannier-function version of the
Berry-phase-based (mean-field) expression for macroscopic polarization
derived by King-Smith and Vanderbilt,\cite{vanderbilt} who gave it
an intuitive interpretation as being a sum over centers of Wannier
functions of the unit cell. Note that expressions above are valid
only in a real-space based approach where square-integrable Wannier
functions are used as single-particle orbitals. If one were to use
Bloch orbitals instead, the expectation value of position operator
will have to be computed differently.\cite{resta} Having demonstrated
the equivalence of our starting expression (Eq. (\ref{eq:corr}))
to the traditional theories at the mean-field level, we next examine
its implications when a many-body expression for the ground state
wave function $|\Psi_{0}^{(\lambda)}\rangle$, expressed in terms
of virtual excitations from the mean-field wave-function, is used\cite{mcweeny}
\begin{eqnarray}
|\Psi_{0}^{(\lambda)}\rangle & = & C_{(0)}^{(\lambda)}|\Phi_{0}^{(\lambda)}\rangle+\sum_{n,\alpha,i,j}C_{n;\alpha}^{(\lambda)}|\Phi_{0}^{(\lambda)}(n\rightarrow\alpha\rangle+\nonumber \\
 &  & \sum_{m,n,\alpha,\beta}C_{m,n;\alpha,\beta}^{(\lambda)}|\Phi_{0}^{(\lambda)}(n\rightarrow\alpha;m\rightarrow\beta)\rangle+\cdots\;,\label{eq:wf}\end{eqnarray}
where the Greek indices $\alpha,\beta,\ldots$ represent the virtual
Wannier functions while the Latin indices $m,n,\ldots$ represent
the occupied ones. $|\Phi_{0}^{(\lambda)}(n\rightarrow\alpha)\rangle$
denotes a singly-excited configuration obtained by promoting one electron
from the occupied Wannier function labeled $n$, to the virtual Wannier
function labeled $\alpha$. Similarly, $|\Phi_{0}^{(\lambda)}(n\rightarrow\alpha;m\rightarrow\beta)\rangle$
represents a doubly-excited configuration with electrons being promoted
from Wannier functions $m,n$ to $\alpha,\beta$. Noteworthy point
is that the occupied ($m,n,\ldots)$ and the virtual Wannier functions
($\alpha,\beta,\ldots)$ could be located in any of the unit cells
of the infinite solid. The coefficients $\{ C_{0}^{(\lambda)},\: C_{n;\alpha}^{(\lambda)},\: C_{m,n;\alpha,\beta}^{(\lambda)},\ldots\}$
can, in principle, be obtained using various available many-body techniques
such as the configurations-interaction (CI) method, perturbation theory,
etc. Next, we examine, the nature of contributions to the polarization
vector ${\bf P^{(\lambda)}}$ arising from virtual excitations when
a correlated wave function ($|\Psi_{0}^{(\lambda)}\rangle$) of the
type of Eq. (\ref{eq:wf}) is used in Eq. (\ref{eq:corr}). In order
to simplify things, we restrict our discussion to the contribution
of the singly-excited configurations $|\Phi_{0}^{(\lambda)}(n\rightarrow\alpha)\rangle$,
although in our calculations all possible excitations needed to compute
both one- and two-body increments have been considered (see discussion
below). Thus, the expectation value of the dipole operator for a singly
excited many-body wave function (assuming that $|\Psi_{0}^{(\lambda)}\rangle$
is real) is\begin{eqnarray}
\langle\Psi_{0}^{(\lambda)}|{\bf R}{}_{e}|\Psi_{0}^{(\lambda)}\rangle &  & =NC_{0}^{(\lambda)2}\langle{\bf r}\rangle_{0}^{(\lambda)}+\nonumber \\
 &  & 2\sqrt{2}N\sum_{n,\alpha}C_{n;\alpha}^{(\lambda)}C_{0}^{(\lambda)}\langle\alpha|{\bf r|}n(o)\rangle^{(\lambda)}+\cdots,\label{eq:sci}\end{eqnarray}
where $\langle{\bf r}\rangle_{0}^{(\lambda)}=\sum_{n=1}^{M}\int{\bf r}|W_{n}({\bf r})|^{2}d{\bf r}$
(the HF expectation value per unit cell), and, using the translational
symmetry, the sum over occupied Wannier functions has been restricted
to those in reference unit cell, denoted as $|n(o)\rangle$, while
the virtual Wannier function $|\alpha\rangle$ can be in any unit
cell of the solid. It is clear from Eq. (\ref{eq:sci}) that: (a)
the expectation value of the dipole operator of the entire solid $\langle\Psi_{0}^{(\lambda)}|{\bf R}_{e}|\Psi_{0}^{(\lambda)}\rangle$
scales linearly with $N$ as it should, and (b) the correlation corrections
to the dipole moment/cell such as the second term of Eq. (\ref{eq:sci}),
can be seen as due to the interactions between the electrons of the
reference unit cell with those in the rest of the solid. Although,
Eq. (\ref{eq:sci}) has been derived for correlated wave functions
containing only singly-excited configurations, however, it is easy
to verify that even for more complex wave functions, only two other
types of matrix elements, viz., $\langle\alpha|{\bf r}|\beta(o)\rangle$
and $\langle m|{\bf r}|n(o)\rangle$ contribute to the correlated
expectation value. These matrix elements originate from interaction
among different types of excited configurations. Because of the localized
nature of the orbitals used, the dipole matrix elements will fall
to zero rapidly with the increasing distance between the orbitals
involved. For example, for a typical insulating solid, the dipole
matrix elements are negligible when the orbitals involved are farther
than third-nearest neighbors. 

Calculation of the correlated many-body wave function (Eq.(\ref{eq:wf}))
of an infinite solid is an extremely difficult task, thereby rendering
the direct use of Eq. (\ref{eq:corr}) even more cumbersome. However,
Stoll\cite{stoll} proposed the use of an {}``incremental'' method
of calculating correlated total energy (and wave function) of extended
systems based upon a Bethe-Goldstone like expansion of correlation
contributions. The approach was subsequently implemented by us to
the case of infinite systems, and utilized to compute the total energy/cell
and related quantities of bulk LiH,\cite{shukla4} and several polymers.\cite{shukla5}
In the approach, the total energy/cell is written as $E_{cell}=E_{HF}+E_{corr}$,
where $E_{HF}$ is the HF energy/cell of the system, and $E_{corr}$
is the contribution of correlation effects to the total energy/cell,
computed as\begin{eqnarray}
E_{corr} & = & \sum_{i}\epsilon_{i}+\frac{1}{2!}\sum_{i\neq j}\Delta\epsilon_{ij}+\nonumber \\
 &  & \frac{1}{3!}\sum_{i\neq j\neq k}\Delta\epsilon_{ijk}+\cdots\label{eq:inc}\end{eqnarray}
where $\epsilon_{i},\:\Delta\epsilon_{ij},\:\Delta\epsilon_{ijk},\ldots$
etc. are respectively the one-, two- and three-body$,\ldots$ correlation
increments obtained by considering simultaneous virtual excitations
from one, two, or three occupied Wannier functions, and $i,\: j,\: k,\ldots$
label the Wannier functions involved.\cite{shukla4} However, using
the incrementally computed many-particle wave function to compute
the expectation value in Eq. (\ref{eq:corr}) is a tedious task which
we avoid by using generalized Hellman-Feynman theorem, and the finite-field
approach to compute dipole expectation values.\cite{mcweeny} Accordingly,
we perform the incremental calculations of energy/cell with the modified
Hamiltonian\begin{eqnarray}
H'({\cal E},\lambda) & = & H_{0}(\lambda)-q_{e}{\cal E}\cdot\sum_{k=1}^{N_{e}}{\bf r}_{k},\label{eq:ffield}\end{eqnarray}
where $H_{0}(\lambda)$ is the usual Born-Oppenheimer Hamiltonian
for the solid with the given value of $\lambda$, and ${\cal E}$
is a user specified external electric field.\cite{nunes} From $E_{cell}^{(\lambda)}({\cal E)}$
so computed, one can easily obtain\begin{eqnarray}
\frac{q_{e}\langle\Psi_{0}^{(\lambda)}|{\bf r}^{j}|\Psi_{0}^{(\lambda)}\rangle}{N} & = & -\frac{\partial E_{cell}^{(\lambda)}({\cal E)}}{\partial{\cal E}_{j}},\label{eq:hellf}\end{eqnarray}
where $j$ represents the Cartesian spatial component ($j=1,2,3)$.
The partial derivatives in Eq. (\ref{eq:hellf}), were computed numerically
by performing the calculation of $E_{cell}^{(\lambda)}({\cal E)}$
for several small values of electric field ${\cal E}$. The Wannier
functions used in the present work were obtained by solving HF equations
in the presence of an electric field, and thus are different from
the ones used in our earlier works.\cite{shukla1,shukla2}

Next we present and discuss our results for the cases of bulk LiH
and LiF. In the present study also we have used the lobe-type contracted
Gaussian basis functions used in our earlier works.\cite{shukla1,shukla2,shukla4,shukla5,shukla6}
Unit cell Wannier functions for both the materials were described
using basis functions centered in cells as far as the third-nearest
neighbors of the reference cell.\cite{shukla1,shukla2} For LiH we
performed the calculations using the optimized lattice constant 4.067
\AA$~$ obtained in our earlier correlated calculation,\cite{shukla4}
which is in excellent agreement with the experimental value of 4.06
\AA. For LiF we used the experimental lattice constant of 3.99 \AA.
For both the systems, fcc geometry was assumed, and anion and cation
locations in the primitive cell were taken to be $(0,0,0)$ and $(a/2,0,0)$,
respectively, where $a$ is the lattice constant. For LiH, in correlated
calculations 1s Wannier function localized on Li as was treated as
core and was held frozen, while for LiF, Wannier functions corresponding
to 1s orbitals of both Li and F were frozen during the correlated
calculations. For computing BEC's, the parameter $\lambda$ corresponds
to atomic displacements $\Delta{\bf u}$ which was taken to be $0.01a$
($a$ is the lattice constant) in the $x$ direction for the anion.
The BEC tensor for cubic materials is diagonal and for the $i$-th
atom of the cell it has only one unique component say $Z^{*}(i)$.
It was computed using the formula $Z^{*}(i)=Z_{i}+(\Omega/e)\frac{\Delta P}{\Delta u}$,
where $Z_{i}$ is the nuclear charge of the displaced atom, $e=|q_{e}|$,
and $\Delta P$ is the change in the unit cell polarization due to
the atomic displacement $\Delta{\bf u}$. For Hellman-Feynman calculations
of the dipole expectation value, we used the central difference formula,
with the values for the external electric field ${\cal E}=\pm0.001$
a.u. in the $x$ direction.

First we verify whether the BEC's computed using the Wannier-function
centers (Eq. (\ref{eq:pol})) agree with those computed using the
Hellman-Feynman theorem (Eq. (\ref{eq:hellf})), at the HF level.
Good agreement between the two calculations at the HF level will be
a vindication of our approach, while any serious disagreement between
the two results will be a setback, and will render further correlated
calculations meaningless. The results from the two calculations are
presented in table \ref{tab-comp} and it is clear that the values
obtained by the two methods are in excellent agreement with each other.
Next we present the results of our correlated calculations in table
\ref{tab-corr}. The table presents the changes in the values of BEC's
as correlation effects of increasing complexity are included using
the aforesaid incremental scheme. The many-body approach used to compute
various correlation increments of Eq. (\ref{eq:inc}) was the full-CI
method, as in our earlier works.\cite{shukla4,shukla5} In case of
LiH we have performed correlated calculation including up to third-nearest
neighbor (3NN) two-body correlation effects,\cite{shukla4} while
for the case of LiF these calculations have been restricted to the
nearest neighbors (NN) two-body increments. The reason behind restricting
the correlation effects for LiF to NN pairs is because the contributions
beyond that (2NN, 3NN, ...) were found to be negligible. This is due
to the fact that for LiH, the hydrogen anion is more diffuse as compared
to the fluorine anion of LiF. Thus the valence electrons of LiH are
comparatively more delocalized as compared to those of LiF, thereby
making the correlation effects relatively longer range in LiH. Inspection
of table \ref{tab-corr} reveals that for the case of LiH, at the
HF level the BEC is overestimated, while for LiF it is underestimated.
When the correlation effects within the reference unit cell are included
for either of the systems, the value of the BEC decreases as compared
to its HF value. This reduction can be seen as due to the mixing of
the occupied Wannier function of anion with the unoccupied ones of
the nearest-neighbor cation, termed ``ion-softening'', by Harrison.\cite{harrison}
However, the noteworthy point is that the ion-softening in the present
case is being driven by the electron-correlation effects. As far as
the longer range correlation effects (1NN, 2NN, ...) are concerned,
no clear trends are visible in table \ref{tab-corr}. For LiH we see
monotonic decrease in the value of BEC with longer range correlation
effects, while for the case of LiF the BEC increases as the nearest-neighbor
correlation effects are included. However, in both the cases, upon
truncation of the correlation series, the values of BEC's obtained
are in excellent agreement with the experimental values. Finally,
in order get a feel for the magnitude of correlation effects with
and without the electric field, we present the values of various contributions
to the correlation energy/cell for the two systems in table \ref{tab-e}
computed for the undistorted unit cell $(\lambda=0)$. Results of
similar calculations performed for the distorted unit cell are not
presented here for the sake of brevity. From the table it is obvious
that, as expected, the most important corrections to $E_{cell}$,
due to the electric field are at the HF level. The correlation energies
in nonzero field are reduced by small amounts as compared to their
zero-field counterparts. Of course, these small changes in correlation
energies for the $\lambda=0$ and $\lambda\neq0$ case in the end
lead to the correlation corrections to the BEC's as depicted in table
\ref{tab-corr}.

In conclusion, we have presented an \emph{ab initio} Wannier-function-based
many-body approach aimed at computing the Born effective charges of
insulators. However, it is clear from the approach that it can also
be used to compute other properties such as high-frequency dielectric
constant, piezoelectric tensor, etc. of insulators. Work along these
lines is presently under way in our group, and the results will be
presented in future publications.

Work of Priya Sony was supported by a Junior Research Fellowship provided
by grant no. SP/S2/M-10/2000 from Department of Science and Technology
(DST), Government of India.%
\begin{table}

\caption{Comparison of Hartree-Fock Born charges of Li computed using the
Wannier center approach (cf. Eq. (\ref{eq:pol})), and via the use
of Hellman-Feynman Theorem (Eq. (\ref{eq:hellf})).}

\begin{tabular}{|c|c|c|}
\hline 
System&
\multicolumn{2}{c|}{Born Charge}\tabularnewline
\cline{2-3} 
&
Wannier-Center Approach&
Hellman-Feynman Approach\tabularnewline
\hline 
LiH&
1.0417&
1.0418\tabularnewline
\hline 
LiF&
0.9986&
0.9983\tabularnewline
\hline
\end{tabular}\label{tab-comp}
\end{table}

\begin{table}

\caption{Influence of electron correlation effects on the Born charge. Column
with heading HF refers to results obtained at the Hartree-Fock level.
Heading {}``one-body'' refers to results obtained after including
the corrections due to {}``one-body'' excitations from each Wannier
function of the unit cell, to the HF value. Two-body (O) implies results
include additional corrections due to simultaneous excitations from
two distinct Wannier functions located on the anion in the reference
unit cell. Two-body (NN), two- body (2NN), and two-body (3NN) correspond
to two-body correlation effects involving 1st, 2nd, and 3rd-nearest
neighboring Wannier functions, respectively. }

\begin{tabular}{|c|c|c|c|c|c|c|c|}
\hline 
System&
\multicolumn{7}{c|}{Born Charge}\tabularnewline
\cline{2-8} 
&
HF&
one-body&
two-body (O)&
two-body (NN)&
two-body (2NN)&
two-body (3NN)&
Exp.\tabularnewline
\hline 
LiH&
1.0418&
1.0302&
---&
1.0193&
1.0183&
1.0003&
0.991$^{a}$\tabularnewline
\hline 
LiF&
0.9983&
0.9913&
0.9847&
1.0237&
---&
---&
1.045$^{b}$\tabularnewline
\hline
\end{tabular}

$^{a}$Brodsky and Burstein \cite{exp-lih} 

$^{b}$Computed from the experimental data reported in Ref. \cite{exp-lif}.\label{tab-corr}
\end{table}
\begin{table}

\caption{Various contributions to energy per unit cell for LiH and LiF in
the absence and the presence of external electric field (${\cal E}$)
for the undisplaced atomic configuration ($\lambda=0.0)$. Similar
calculations were also performed for distorted unit cells. For the
sake of brevity we are displaying the correlation contributions only
up to the NN two-body terms. But $E_{cell}$ includes all the computed
contributions.}

\begin{tabular}{|c|c|c|c|c|c|}
\hline 
System&
${\cal E}$&
\multicolumn{4}{c|}{Energy per cell (a.u.)}\tabularnewline
\cline{3-6} 
&
(a.u.)&
$E_{HF}$&
$E_{corr}(\mbox{O})$&
$E_{corr}(\mbox{NN})$&
$E_{cell}$\tabularnewline
\hline 
LiH&
0.000&
-8.061995&
-0.029551 &
-0.004153&
-8.096270\tabularnewline
&
0.001&
-8.065846&
-0.029549&
-0.004153&
-8.100117\tabularnewline
\hline 
LiF&
0.000&
-107.045079&
-0.144748&
-0.004434&
-107.194261\tabularnewline
&
0.001&
-107.048856&
-0.144744&
-0.004430&
-107.198030\tabularnewline
\hline
\end{tabular}\label{tab-e}
\end{table}


\begin{thebibliography}{10}
\bibitem{born}M. Born and K. Huang, \emph{Dynamical Theory of Crystal Lattices}
(Clarendon, Oxford, England, 1954).
\bibitem{zhong}W. Zhong, R.D. King-Smith, and D. Vanderbilt, Phys. Rev. Lett. \textbf{72},
3618 (1994).
\bibitem{car}See, \emph{e.g.}, A. Pasquarello and R. Car, Phys. Rev. Lett. \textbf{79},
1766 (1997).
\bibitem{vanderbilt}R.D. King-Smith and D. Vanderbilt, Phys. Rev. B \textbf{47}, 1651
(1993).
\bibitem{dall}S. Dall'Olio, R. Dovesi, and R. Resta, Phys. Rev. B \textbf{56}, 10105
(1997).
\bibitem{martin-1}G. Ortiz and R.M. Martin, Phys. Rev. B \textbf{49}, 14202 (1994).
\bibitem{martin-2}I Souza, T. Wilkens, and R.M. Martin, Phys. Rev. B \textbf{62}, 1666
(2000).
\bibitem{filip}A. Filippetti and N.A. Spaldin, Phys. Rev. B. \textbf{68}, 045111
(2003).
\bibitem{shukla1}A. Shukla, M. Dolg, H. Stoll, and P. Fulde, Chem. Phys. Lett. \textbf{262},
213 (1996).
\bibitem{shukla2}A. Shukla, M. Dolg, P. Fulde, and H. Stoll, Phys. Rev. B \textbf{57},
1471 (1998).
\bibitem{shukla4}A. Shukla, M. Dolg, P. Fulde, and H. Stoll, Phys. Rev. B \textbf{60},
5211 (1999). 
\bibitem{shukla5}A. Abdurahman, A. Shukla, and M. Dolg, J. Chem. Phys. \textbf{112},
4801 (2000).
\bibitem{shukla6}A. Shukla, Phys. Rev. B \textbf{61}, 13277 (2000).
\bibitem{note}This formula can be seen as the real-space counterpart of the ${\bf k}$-space-based
Berry-phase many-body expression for macroscopic polarization of bulk
insulators, obtained by Souza, Wilkens, and Martin (see Eqs. (12)
and (38) of ref. \cite{martin-2}).
\bibitem{mcweeny}See, e.g., R. McWeeny, \emph{Methods of Molecular Quantum Mechanics},
2nd edition (Academic Press, London, 1989).
\bibitem{resta}R. Resta, Phys. Rev. Lett. \textbf{80}, 1800 (1998).
\bibitem{stoll}H. Stoll, Phys. Rev. B \textbf{46}, 6700 (1992); H. Stoll, Chem. Phys.
Lett. \textbf{191}, 548 (1992).
\bibitem{exp-lih}M.H. Brodsky and E. Burstein, J. Phys. Chem. Solids \textbf{28}, 1655
(1967).
\bibitem{exp-lif}M.J.L. Sangster, U. Schr\"oder, and R.M. Atwood, J. Phys. C. \textbf{11},
1523 (1978).
\bibitem{nunes}See, \emph{e.g.}, R.W. Nunes and D. Vanderbilt, Phys. Rev. Lett. \textbf{73},
712 (1994).
\bibitem{harrison}W. Harrison, \emph{Electronic Structure and Properties of Solids}
(Freeman, San Francisco, 1980).
\end{thebibliography}
\end{document}